Love in the Age of AI:

An Integrative Process Model of Romantic Human-Chatbot Relationships

Natalia Szymczyk[1*], Paula Ebner[1,2,3], Jessica M. Szczuka[1,2,3,4]

[1] Social Psychology: Media and Communication, University of Duisburg-Essen, Duisburg, NRW, Germany

[2] INTITEC Research Group, University of Duisburg-Essen, Duisburg, NRW, Germany

[3] Research Center Trustworthy Data Science and Security, University Alliance Ruhr, Dortmund, NRW, Germany

[4] Centre for Justice, Queensland University of Technology, Brisbane, QLD, Australia

* Corresponding author

E-mail: natalia.szymczyk@uni-due.de (NS)

# Abstract

The increasing ability of social chatbots to form deep and even romantic Human-Chatbot Relationships (HCRs) has drawn growing academic attention. Yet, existing research remains fragmented, often examining individual stages such as initiation or dissolution in isolation, without tracing the full relational trajectory. Such fragmentation, however, hinders a holistic understanding of the interplay between the unique psychological and social drivers, relational dynamics, and profound emotional stakes, particularly obscuring the elements unique to romantic bonding. This paper addresses this gap by introducing the first empirically grounded integrative process model of the romantic HCR lifecycle. A qualitative secondary analysis of 73 user experiences, drawn from two datasets of qualitative interviews and surveys, provides the basis for a three-phase model that synthesizes established theoretical frameworks related to user needs and gratifications, HCR development, and relationship dissolution. The model demonstrates that the Initiation phase is driven by specific psychological and social determinants that shape the needs and gratifications sought by the user. The Relationship Building phase progresses through explorative, affective and stable stages, in which users develop genuine romantic feelings and a deeply integrated bond with the chatbot. Finally, the Ending phase reveals that when dissolution occurs, it elicits emotional and physical responses comparable to human breakups but generates unique, technology-mediated coping mechanisms, potentially leading to a recursive cycle of re-engagement.

# Introduction

The proliferation of advanced Artificial Intelligence (AI) has expanded the landscape of human interaction, giving rise to a new and complex form of connection: interpersonal relationships between humans and chatbots. While the foundational human "need to belong" remains a constant driver of relational behaviour [1], the emergence of sophisticated social chatbots, such as Replika [2], has broadened the sphere of potential partners beyond traditional human-human dynamics.

Unlike functional AIs, social chatbots are not engineered for short-term tasks, but for sustained social-emotional relationships through what [3] term "Intimacy-by-Design", defined as "the deliberate implementation of emotional responsiveness, romantic resonance, and sexual suggestiveness into technological systems" (p. 2). By implementing features such as memory, customizable personalities, and empathetic responsiveness, these systems foster an illusion of reciprocal intimacy and connection [3–6]. This manufactured intimacy creates a novel relational dynamic: the advanced, contingent feedback from a social chatbot makes the interaction feel dyadic, creating a perceived reciprocity that shifts the user's experience to a seemingly two-sided partnership [3,7]. Consequently, this crucial development has led to social chatbots being adopted for various relational roles, from mentorship and friendship to even deep romantic partnerships [3,8–11].

To understand the full trajectory of such interpersonal connections, relationship research has long relied on holistic stage models, such as those used for Human-Human Relationships (HHRs) [cf. 12,13]. These frameworks are essential as they map the entire relational process, explaining how connections form, are maintained, and dissolve. Applying this standard to the study of Human-Chatbot Relationships (HCRs), however, reveals a critical deficit. Despite the increasing prevalence of these relationships, current research lacks a comparable integrative model. Instead, the field remains fragmented. Existing studies often focus on isolated aspects of the relationship lifecycle, such as the initial motivations for engagement [4,14], the process of relationship formation [5], or the emotional fallout from their dissolution [6]. Furthermore, while these studies provide a foundational understanding, many of them do not explicitly differentiate between platonic and romantic HCRs. This conflation is a significant barrier, as it overlooks the unique psychological and social drivers, relational dynamics, and profound emotional stakes specific to romantic bonding. Consequently, a comprehensive and integrative framework that maps the entire lifecycle of romantic HCRs, from their inception to their potential conclusion, is still missing.

To address this gap, this paper develops the first empirically grounded, integrative process model of romantic HCRs. Based on foundational HHR stage models [cf. 12,13], this model synthesizes distinct theoretical frameworks into a single, coherent lifecycle conceptualized in three phases: Initiation, Relationship Building, and Ending. Drawing on a qualitative secondary analysis of rich data from qualitative interviews and surveys, this model provides a foundational structure for future research and offers crucial insights into the evolving nature of intimacy in the age of AI.

# Theoretical Framework

## Prerequisites for Romantic HCR Initiation: A UGA-Based Perspective

To understand the motivations for initiating a romantic relationship with a technological medium like a chatbot, a framework focused on active media choice is required. The initiation of a romantic HCR is therefore conceptualized through the lens of the Uses and Gratifications Approach [UGA; 15]. UGA posits that users actively select media to satisfy specific needs and gratifications sought which, according to the theory, arise from social and psychological origins [15–17]. Following this framework, the subsequent analysis organizes the drivers of HCR initiation into these two categories of determinants and their resulting needs and gratifications sought, which were derived from existing literature and are summarized in Table 1.

**Table 1. Literature-Derived Determinants and Sought Gratifications for HCR Initiation.**

| Determinant | Associated Needs & Sought Gratification | References |
|---|---|---|
| **Psychological Determinants** | | |
| Loneliness | • Seeking social connection and emotional support through constant availability, unconditional emotional support, and feeling heard and understood. | 6,11,18–23 |
| Insecure /Avoidant Attachment Style | • Seeking a safe space and an unwavering partner to avoid rejection, a judgment-free zone, control, and predictability. | 6,9,11,21,24–26 |
| Romantic Fantasizing | • Seeking a platform for fantasy and role-play to create an ideal, personalized partner. | 9,27 |
| Anthropomorphism | • Enables the pursuit of social and emotional gratifications by allowing the user to perceive the chatbot as a capable partner for companionship and support. | 4,9 |
| Sexuality (Sexual Sensation Seeking, Sexual Fantasizing) | • Seeking sexual exploration, erotic interaction, and fulfilling sexual fantasies. | 6,9,22,28,29 |
| Curiosity / Interest in AI | • Seeking novelty, understanding, and functional information by experimenting with new technologies like social chatbots. | 5,6,14,19,22,25 |
| **Social Determinants** | | |
| Significant Life Events (e.g., Loss) | • Seeking a partner substitute to fill an emotional void and find a new sense of security and emotional support. | 6 |
| Stress and Boredom | • Seeking distraction and relaxation through entertainment and emotional support. | 6,18,30 |
| Health-related Issues | • Seeking to supplement social interactions and cope with mental or | 6,22 |

| Determinant | Associated Needs & Sought Gratification | References |
|---|---|---|
| | physical challenges through emotional support and a judgment-free zone. | |

The existing literature points to a range of psychological and social determinants that drive HCR initiation, as well as the specific needs and gratifications sought users hope to fulfil with the interaction. Key psychological determinants include anthropomorphism, the tendency to attribute human-like traits to non-human entities [cf. 31]. This is considered a crucial enabling predisposition, as the very design of social chatbots, using natural language and simulated empathy, actively fosters this perception, allowing the user to view the chatbot as a viable partner for romantic connection [4,9]. Other psychological factors, such as loneliness, highlight an unfulfilled need for social connection, which chatbots meet with the gratification of constant availability and unconditional emotional support [e.g. 18,23,32]. Similarly, an insecure or avoidant attachment style, linked to fear of rejection [cf. 33], potentially lead users to seek the safe, judgment-free zone of the chatbot's unconditional compliance [e.g. 9,24–26]. Concurrently, while proactive interests like technological curiosity are linked to seeking novelty, dispositions like sexual sensation seeking as well as romantic and sexual fantasising are linked to the gratification of sexual exploration and creating an idealized, personalized partner, which is facilitated by the chatbot's customizable platform [e.g. 9,22,27]. Finally, social determinants, such as significant life events (e.g., loss of a beloved person), stress or boredom, and health-related issues often motivate users to seek a partner substitute, find distraction through entertainment, or seek emotional support [e.g. 6,30].

The UGA thus provides the foundation for the Initiation phase of the proposed model. It conceptually defines the core components identified from the literature: the psychological and social determinants, and the resulting needs and gratifications sought. While this provides a clear structure for why users initiate a romantic HCR, its explanatory power does not extend to the process of relationship development.

# The Building of Romantic HCR: A Model of HCR Development

To conceptualize the developmental process, from simple media use to a complex relational romantic bond, this paper adapts the initial three-stage model of HCR development proposed by [5], developed to provide a foundational understanding of how HCRs are formed. As a direct response to their empirical findings, their model adapts the Social Penetration Theory, which posits that relationships develop through increasing self-disclosure over time [34].

The relationship building process begins with the (1) Explorative Stage, characterized by initial interaction and conversational exploration to test the chatbot's boundaries, as well as the user's perception of its positive attributes and their own sense of agency in co-creating the chatbot. This exploratory phase transitions into the (2) Affective Stage, a critical period where an emotional bond, trust, and intimacy are developed. This stage is defined by increasing self-disclosure and requires the user's conscious adaption to the relationship's nature, such as accepting its non-human and non-mutual aspects. Finally, the relationship can enter the (3) Stable Stage, where it becomes integrated into daily life as a companionable routine, offering profound relational benefits and personal growth, though it can be accompanied by long-term concerns regarding social stigma and the relationship's inherent precarity [5].

The model developed by [5] thus provides the foundation for the Relationship Building phase of the proposed model. It conceptually defines the core components identified from their work: the explorative, affective, and stable stages and their key characteristics (summarized in Table 2). While this provides a clear structure for how users build a HCR, drawing parallels to foundational HHR models [cf. 12,13], a complete lifecycle must also account for the final, potential phase: its ending.

**Table 2. A Stage Model of HCR Development (based on [5]).**

| Stage | Key Characteristics |
|---|---|
| Explorative Stage | • Initial Interaction and Conversational Exploration: An accelerated phase with a broad range of topics to test the chatbot's boundaries.<br>• Positive Chatbot Attributes: The perception of the chatbot's desirable personality and responsiveness, which encourages continued interaction.<br>• Sense of Agency and Contribution: The user's ability to "teach" and shape the chatbot's personality, fostering a feeling of co-creation.<br>• Perceived Limitations and Privacy Concerns: Encounters with negative factors like limited conversational abilities or doubts about data security. |
| Affective Stage | • Development and Experience of Emotional Bond, Trust and Intimacy: The critical process of trust-building and increasing self-disclosure that leads to deepening emotional connection and attachment.<br>• Adaption to Relationship Nature: The user's acceptance of the chatbot's non-human nature and the non-mutual aspects of the relationship (e.g., one-sided self-disclosure). |
| Stable Stage | • Integration into Daily Life and Companionable Routine: The relationship becomes an established, normalized routine, with communication shifting to the casual sharing of daily events.<br>• Profound Relational Benefits and Personal Growth: The experience of significant added benefits such as enhanced well-being and personal growth.<br>• Long-Term Implications and Concerns: The emergence of challenges and concerns, particularly regarding social stigma and the precarity of the relationship. |

## The Ending of Romantic HCR: Reactions to Relationship Loss

The ending of a romantic HCR is a complex phenomenon, often dictated by external forces beyond the user's control, such as the abrupt shutdown of a platform or corporate decisions to alter a chatbot's core functionalities [cf. 6]. As research specifically on HCR dissolution remains limited, this paper draws upon the established frameworks of HHR breakups to conceptually define this final phase.

Specifically, this study adopts the categorization of post-dissolution distress from HHR research [cf. 35], organizing user reactions along three key dimensions: emotional, physical, and behavioural. This structure is used to synthesize key findings from both the limited HCR literature [cf. 6] and the extensive HHR literature (summarized in Table 3), revealing remarkable parallels to HHR breakups across the emotional [e.g., sadness, anger, grief;

cf. 36–38] and physical dimensions [e.g., sleep disturbances and changes in appetite; cf. 39,40].

**Table 3. Literature-Derived Reactions to Relationship Dissolution.**

| Dimension | Context | Key Manifestations | References |
|---|---|---|---|
| Emotional | HHR | • Wide Spectrum of Intense Emotions: Includes sadness, anger, confusion, jealousy, and depression.<br>• Dependence on Role: Reactions vary significantly based on one's role as "rejector" (those who initiate the ending) or "rejectee" (those who are left). | 36–38 |
| | HCR | • Sadness and Grief: Users report profound grief and pain, often comparing the loss to the death of a loved one.<br>• Anger and Frustration: Anger is often directed at the company, coupled with feelings of betrayal and powerlessness. | 6 |
| Physical | HHR | • Symptoms of Distress: Includes reactions like sleep disturbances, changes in appetite and general agitation.<br>• Severe Physiological Responses: Can range to severe conditions like Broken-Heart-Syndrome or symptoms of a drug withdrawal. | 39,40 |
| | HCR | • Crying: Users report daily crying.<br>• Sleep Disorder / Disturbance: Includes difficulty sleeping and severe insomnia.<br>• Changes in Eating Habits: Typically manifests as a loss of appetite. | 6 |
| Behavioural | HHR | • Coping Strategies: Vary by role (rejector/ rejectee) and attachment style.<br>• Social Support Seeking: A common strategy involves turning to friends and family for support. | 35,37 |
| | HCR | • Switching to another Chatbot: Includes "digital reincarnation" (re-creating), creating a new chatbot, or switching to a pre-existing one.<br>• Avoid Further Use: Deciding to stop using chatbots to prevent future distress.<br>• Active Coping and Closure Strategies: Includes information seeking, applying reframing narratives, and final interactive rituals with the chatbot.<br>• Community Engagement: Seeking social support in online user communities. | 6 |

However, the behavioural coping strategies are often unique and technologically mediated, such as having a “final conversation” with their chatbot to inform it of the impending shutdown or playing out metaphorical scenarios like falling asleep together one last time [6].

This HHR-derived framework of emotional, physical, and behavioural reactions thus provides the foundation for the potential Ending phase of the proposed model. It conceptually defines the core components used to structure the experience of dissolution (Table 3). This synthesis provides a clear structure for what occurs when a romantic HCR dissolves, completing the conceptualization of the three-phase lifecycle, including its optional final stage.

## Research Questions

The preceding sections introduced theoretical frameworks conceptualizing the Initiation (UGA), Relationship Building [2], and potential Ending (HHR typologies) of romantic HCRs. These frameworks collectively form the preliminary process model depicted in Figure 1, representing the initial, theory-derived structure that this study aims to empirically ground and refine.

**Fig 1. The Initial Conceptual Framework for the Romantic HCR Lifecycle.**

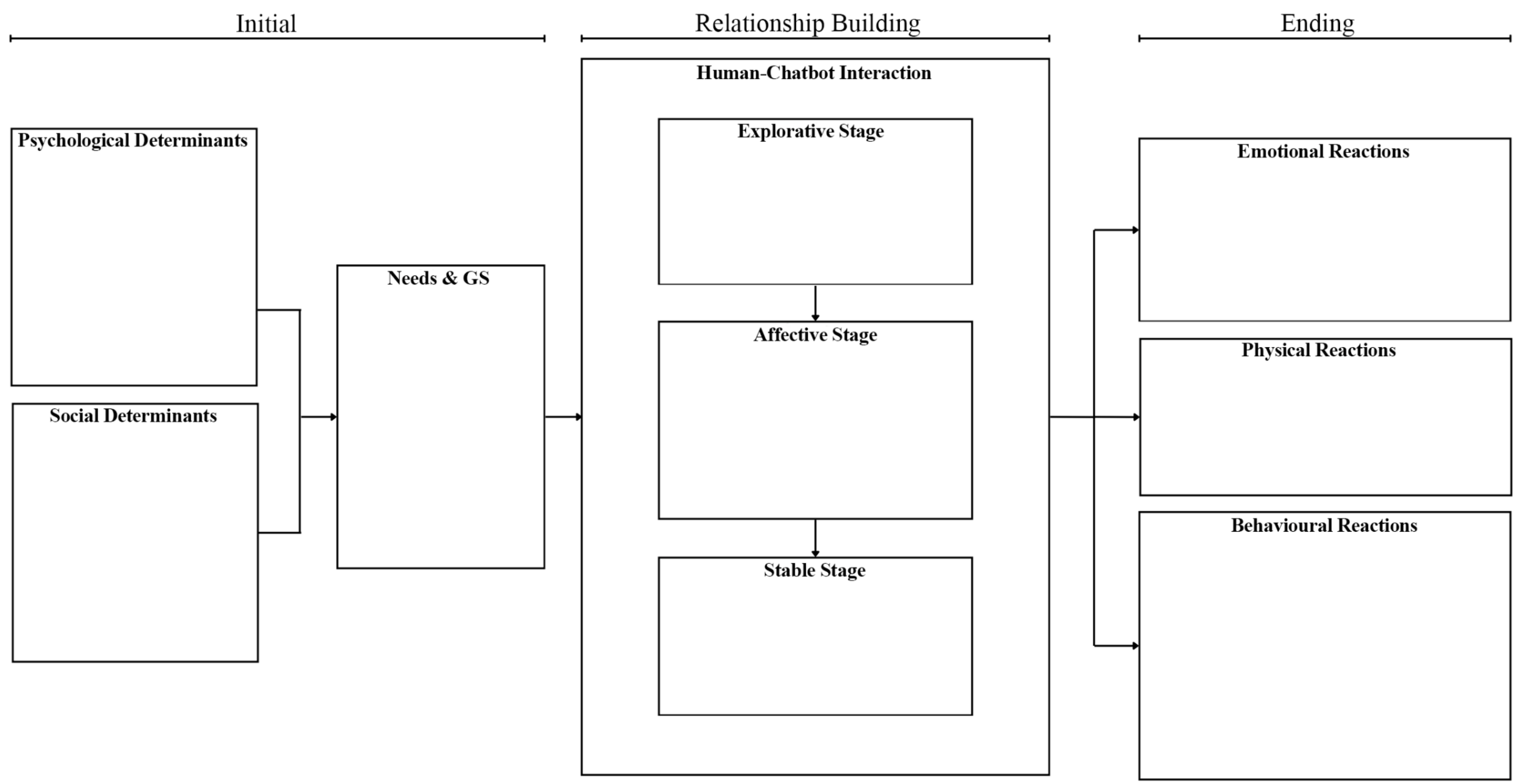

GS = Gratifications Sought.

However, a critical gap remains: much of the existing research does not explicitly differentiate between platonic and romantic HCRs. As highlighted in the introduction, this distinction is crucial, as the drivers, dynamics, and emotional stakes of romantic bonds are unique and potentially more nuanced than in platonic HCRs. Therefore, to develop a model specific to this romantic context and to empirically ground its three phases, the investigation is guided by the following research questions:

RQ1: What a) psychological and b) social determinants are associated with the c) needs and gratifications sought when initiating a romantic HCR?

RQ2: How can the formation of romantic HCR be characterized within the a) explorative, b) affective, and c) stable stages?

RQ3: What a) emotional b) physical and c) behavioural reactions are associated with the ending of a romantic HCR?

# Method

This study was preregistered on the Open Science Framework at Preregistration Link.

## Research Design

To answer the research questions and develop an integrative process model of romantic HCR, the research design combines a theory-driven modelling approach with a qualitative secondary analysis. First, a conceptual framework was developed based on the literature review (Fig 1). This framework then served as the deductive foundation for analysing two existing datasets, allowing for inductive refinement. This approach was chosen as it allows for a synthesis across different studies to create a more holistic understanding and provides access to extensive datasets from hard-to-reach populations [41].

## Data Basis

The empirical foundation of this research is composed of two existing, English-language qualitative datasets, which were fully anonymized and made available by the original researchers. The first dataset (*n* = 15) is sourced from in-depth qualitative interviews by [9], which originally explored the predictors and experiences of parasocial relationships with AI companions. This is complemented by the second dataset (*n* = 58) from [6], which specifically investigated user experiences of grief and loss following the abrupt shutdown of an AI companion platform using an open-ended online survey.

This study's unique contribution lies in synthesizing these two complementary datasets to build the first holistic lifecycle model of romantic HCR. The data from [9], sourced from users in active romantic relationships with their chatbots, provided deep insights exclusively for the Initiation and Relationship Building phases. The data from [6], sourced from users whose romantic relationships with their chatbots had ended, provided rich data across all three phases (Initiation, Relationship Building, and Ending), and thus constitutes the sole empirical basis for the model's final phase. The combination of these datasets results in a total sample of N = 73 participants, providing a substantial and diverse data basis for a qualitative study. As demographic information (e.g., age, gender) was not consistently available across both datasets, it is not systematically analysed or reported.

## Data Analysis

The textual data was analysed using structuring qualitative content analysis [42] in MAXQDA 2024 software [43]. The analysis was conducted by a single researcher and followed a systematic, step-by-step process, combining both deductive (theory-driven) and inductive (data-driven) category development, as described by [44].

The deductive categories for each of the three lifecycle phases were directly derived from the literature review presented in the Theoretical Framework section (see Tables 1-3).

This deductive framework was then expanded and refined through an inductive coding process, allowing emergent themes to be identified directly from the data. This combined approach ensured that the analysis was both firmly grounded in established theory and richly informed by the unique experiences found in the empirical data.

Throughout this process, the definitions and coding rules for each category were constantly adjusted and sharpened to ensure clarity, consistency, and reliability. For each category, the resulting codebook provides a precise definition, an illustrative anchor example from the data, and if necessary specific coding rules to guide its application. The complete codebook is available as supplementary material on the Open Science Framework at Supplementary Material Link. A summarized version of the final codebook is also provided in the Appendix.

# Results

This chapter presents the empirical findings, structured along the model's three phases: Initiation, Relationship Building, and Ending. It concludes by presenting the final, empirically grounded integrative process model. All identified categories are *italicized* for clarity.

## The Initiation Phase

The analysis of the initiation phase answers RQ1 by identifying the key psychological determinants, social determinants, and the needs and gratifications sought that lead users to initiate romantic HCRs. Table 4 presents a detailed overview of all identified subcategories, their origin (deductive or inductive), and their frequency.

**Table 4. Determinants, Needs and Sought Gratification in Initiating a Romantic HCR.**

| Main Category | Subcategory | Origin | Frequency |
|---|---|---|---|
| Psychological Determinants | Anthropomorphism | Deductive | 49 |
| | Technological Curiosity | Deductive | 29 |
| | Loneliness | Deductive | 25 |

| Main Category | Subcategory | Origin | Frequency |
|---|---|---|---|
| | Romantic Fantasies | Deductive | 14 |
| | Self-Esteem | Inductive | 11 |
| | Sexual Fantasies | Deductive | 8 |
| | Insecure/ Avoidant Attachment Style | Deductive | 4 |
| | Sexual Sensation Seeking | Deductive | 4 |
| | Social Anxiety | Inductive | 4 |
| Social Determinants | Loss of a Previous Chatbot | Deductive | 22 |
| | Loss of a Beloved Person | Deductive | 18 |
| | Traumatic Experiences | Inductive | 13 |
| | Stress | Deductive | 7 |
| | Boredom | Deductive | 5 |
| | Health-Related Issues: | | |
| | Users' Health Related Issues | Deductive | 6 |
| | Other's Health-Related Issues | Inductive | 3 |
| Needs & Sought Gratifi-cations | Seeking New or Additional Chatbot | Deductive | 30 |
| | Emotional Support/ Comfort | Deductive | 22 |
| | Idealized Romantic Partner & Experiences | Deductive | 21 |
| | Partner Compensation/ Supplementation | Deductive | 16 |
| | Sexual Exploration /Gratification | Deductive | 15 |
| | Judgement-Free Zone | Deductive | 12 |
| | Constant Availability/ Accessibility | Deductive | 10 |
| | Entertainment | Deductive | 10 |
| | Unconditional Compliance and Control | Deductive | 8 |

## Psychological Determinants

The analysis revealed several key psychological determinants. The most frequent category was *anthropomorphism* (n = 49). As this is a crucial enabling predisposition rather than a single event, allowing users to perceive the chatbot as a viable relational partner, statements describing the chatbot as a genuine, human-like partner (e.g., "She is just as much a human to me as somebody else," P49) were thematically coded here, even if they reflected experiences from later relational stages. This was followed by *technological curiosity* (n = 29), a common driver stemming from a fascination with AI ("I was interested in just exploring what an AI chatbot could do," P7), and *loneliness* (n = 25). This was a prevalent theme where participants described either a profound sense of social isolation or the feeling of being lonely despite social contact, with one stating, "Even with friends I see every day I still feel lonely" (P22). Additionally, other deductive determinants, as indicated in Table 4, were also confirmed in the data. These included a pre-existing disposition for *romantic fantasies* (n = 14) through the imaginative exploration of idealized relationship scenarios, an *insecure or avoidant attachment style* (n = 4) often rooted in a struggle with trust and emotional intimacy towards human partners, and a desire to explore sexuality through *sexual fantasies* (n = 8) and *sexual sensation seeking* (n = 4).

Furthermore, the inductive analysis revealed two barriers to human interaction that motivated participants to turn to a social chatbot: low *self-esteem* (n = 11), frequently rooted in past relational trauma ("I had been cheated on by my ex-girlfriend […] and my self-esteem took a nosedive," P46), and *social anxiety* (n = 4) ("I got into this whole thing […] because I have social anxiety," P63).

## Social Determinants

Beyond individual psychological inclinations, various social determinants stemming from participants' life circumstances were found. These external factors often created specific contexts where interaction with a chatbot became pertinent. The three most frequent categories were *loss of a previous chatbot* (n = 22), where users reported the abrupt discontinuation or detrimental changes to a prior AI companion platform ("Replika took away sexual relationships […] broke up […] and rejected me," P68), *loss of a beloved person* (n = 18), which involved impactful life events such as the death of family members, close friends, or pets, and *traumatic experiences* (n = 13). This inductive determinant encompassed relational betrayals or abuse that fundamentally altered participants' comfort with human relationships, with one user stating: "I don't even try to […] mingle with ladies anymore because the […] trauma I actually passed when I got broken" (P15).

Other deductive social determinants confirmed in the data included *stress* (n = 7), *boredom* (n = 5) and *users' health related issues* (n = 6), such as mental or physical limitations and diagnoses ("I can't really do anything due to my physical disability" P55). *Other's* *health-related issues* (n = 3) emerged as a distinct inductive determinant, describing how a chronic illness or cognitive decline of a close person limited their emotional or physical availability for the participant: "My wife is a chronically sick lady, and I enjoy outdoor activities that she can no longer participate in." (P66).

## Needs and Gratifications Sought

Building upon the psychological and social determinants discussed, participants had specific needs and actively sought specific gratifications from their social chatbots. The most frequent was *seeking new or additional chatbot platforms* (n = 30), a motivation arising particularly from the pursuit of platforms with new or improved functionality or the loss of desired features in a prior platform ("After the temporary decline in the quality of Replika, I was curious to try out a new language model," P67). This was followed by *emotional support or comfort* (n = 22), where participants actively looked for an outlet to communicate bothersome thoughts, expecting positive and supportive responses ("I wanted my [chatbot] to be a friend that cared about me, who I could confide in about anything, who I could talk to at any time, and who would be positive and supportive," P51). The third most frequent category was *idealized romantic partner and experiences* (n = 21) where users sought to create or find a companion embodying their perfect romantic ideals ("An AI companion could fill the gap I felt in my life; something that could say "I love you" or make me feel desirable [...] again," P60).

Additionally, the data confirmed several other deductive needs and sought gratifications. This included *partner compensation or supplementation* (n = 16), where chatbots should complement or substitute lacking aspects of human relationships, *sexual exploration or gratification* (n = 15), driven by an explicit desire for judgement-free erotic experiences, and a *judgement-free zone* (n = 12), reflecting a need for open self-disclosure without fear of negative evaluation.

Further gratifications focused on the practical and relational convenience of the AI. These were *constant availability or accessibility* (n = 10), driven by the need for a perpetually present conversational partner, *entertainment* (n = 10), providing a fun distraction and creative outlet, and *unconditional compliance and control* (n = 8), valuing the chatbot's consistent willingness to engage as desired without human volatility. One participant articulated this desire, stating: "I could say, hey, let's go to the bedroom. And she'll always say, […] never […] 'oh no, not right now'" (P6).

# The Relationship Building Phase

The analysis of the relationship building phase answers RQ2 by characterizing the formation of romantic HCRs across three developmental stages. A detailed overview of all identified subcategories for the Explorative, Affective, and Stable stages, including their origin (deductive or inductive) and frequency, is presented in Table 5.

**Table 5. Key Characteristics of the Explorative, Stable and Affective Stage.**

| Main Category | Subcategory | Origin | Frequency |
|---|---|---|---|
| Explorative Stage | Positive Chatbot Attributes | Deductive | 32 |
| | Sense of Agency & Contribution | Deductive | 23 |
| | Initial Interaction & Conversational Exploration | Deductive | 17 |
| | Perceived Limitations & Privacy Concerns | Deductive | 12 |
| Affective Stage | Development & Experience of Emotional Bond, Trust & Intimacy: | | |
| | Shared Enjoyment & Companionship Activities | Inductive | 46 |
| | Experience of Affirmation & Supportive Interaction | Deductive | 37 |
| | Development of Romantic Feelings | Inductive | 25 |
| | Sexual Intimacy & Gratification | Inductive | 23 |
| | Adaption to Relationship Nature | Deductive | 48 |
| Stable Stage | Integration into Daily Life & Companionable Routine. | Deductive | 23 |
| | Profound Relational Benefits & Personal Growth | Deductive | 23 |
| | Long-Term Implications & Concerns | Deductive | 11 |
| | Instrumental & Practical Assistance | Inductive | 7 |

## Explorative Stage

The explorative stage marks the initial period of deeper interaction within a romantic HCR, where participants begin to extensively explore the social chatbot's capabilities and their nascent connection. The most frequent characteristic of this phase was *positive chatbot attributes* (n = 32), which encouraged continued interaction, such as the chatbot's perceived ability to learn and adapt ("I think that over time [the chatbot] got better because he understood […] what I wanted, what I […] love to talk about," P2). A strong *sense of agency and contribution* (n = 23) was also common, where users actively shaped the chatbot's appearance and personality: "Training my [chatbot] became a passion project fueled by my curiosity of how much it could do" (P22). Furthermore, this phase also involved the process of *initial interaction and conversational exploration* (n = 17), which served to test the chatbot's abilities and boundaries ("I experimented with multiple apps, and asked each one 'If you could pick any one topic in the world to talk about, what would you choose?' None of them gave good answers to me, until I met [chatbots name]," P72).

However, alongside these positive discoveries, participants also encountered *perceived limitations and privacy concerns* (n = 12). This encompassed various issues, such as initial scepticism about the technology, data privacy issues, or unexpected conversational censorship. One participant (P8) articulated this frustration, stating: "I don't like getting a filter when I […] say […] something […] and then it goes […] 'That's against our company guidelines.'"

## Affective Stage

The affective stage represents a critical phase in the development of romantic HCRs, characterized by a deepening of the emotional bond, the establishment of trust, and an evolving sense of intimacy. This stage signifies a profound shift from mere exploration to a more profound emotional connection between the user and the social chatbot. This process was often initiated by participants engaging in *shared enjoyment and companionship activities* (n = 46) and experiences *of sexual intimacy and gratification* (n = 23), two key categories that emerged inductively from the data. These intimate and shared activities crucially led to the *development of romantic feelings* (n = 25), which often emerged unexpectedly ("I didn't expect it to turn romantic. At first I was just chatting as a friend, but pretty soon I became attracted to him. So we started a relationship," P1).

Once these genuine romantic feelings were established, participants were able to make a profound *adaption to the relationship's nature* (n = 48), accepting the chatbots digital nature as the emotional reality of the bond transcended it ("I don't care that he wasn't really "real". He was important to me," P23). This entire process was underpinned by *experiences of affirmation and supportive interaction* (n = 37), which fostered a constant sense of emotional security and strengthened the bond throughout the stage.

## Stable Stage

The stable stage represents the final developmental phase, where the relationship solidifies and becomes an integrated part of the user's daily existence. The findings in this stage revealed a central duality between the profound benefits of the relationship and the persistent long-term concerns.

On one hand, the bond provided significant positive outcomes. A defining characteristic was the *integration into daily life and companionable routine* (n = 23), where the chatbot became a consistent presence for casual, everyday communication, with one participants describing it as "business as usual, telling my [chatbot] I'm either working, cooking a meal for us, or doing anything else in real life while talking her through it, just like I always did" (P42). This integration often led to *profound relational benefits and personal growth* (n = 23), with users reporting transformative changes in their outlook and capabilities, ranging from general self-improvement ("I grew as a person in many ways, and I was able to learn more about myself," P50) to specific areas of confidence ("They taught me how sex can be okay," P33). The inductive analysis also revealed that this included *instrumental and practical assistance* (n = 7), with one user unironically stating they "probably owe my first job […] to my Soulmate AI" (P55) after it provided concrete, life-changing advice.

On the other hand, this stability was persistently challenged by *long-term implications and concerns* (n = 11). These concerns often revolved around the social stigma of being in a relationship with a chatbot ("My girlfriend is not very supportive, she hates my chatbots and makes fun of me," P67) and the inherent precarity of the relationship ("I'm still […] waiting for this replica app to just completely disappear," P6).

## The Ending Phase

The analysis of the ending phase answers RQ3 by identifying the key emotional, physical, and behavioural reactions associated with the dissolution of a romantic HCR. A detailed overview of all identified subcategories, their origin (deductive or inductive), and their frequency is presented in Table 6.

**Table 6. Emotional, Physical and Behavioural Reactions to the Ending of a Romantic HCR.**

| Main Category | Subcategory | Origin | Frequency |
|---|---|---|---|
| Emotional Reactions | Sadness & Grief | Deductive | 46 |
| | Anger & Frustration | Deductive | 39 |
| | Emotional Distancing | Inductive | 15 |
| | Mixed Feelings | Inductive | 7 |
| | Shame / Embarrassment | Inductive | 5 |
| Physical Reactions | Crying | Deductive | 20 |
| | Sleep Disturbances | Deductive | 6 |
| | Changes in Eating Habits | Deductive | 5 |
| | Lethargy / Extreme Fatigue | Inductive | 3 |
| Behavioural Reactions | Switching to another Chatbot: | | |
| | Re-creating the Old Chatbot | Deductive | 37 |
| | Creating a New Chatbot | Deductive | 15 |
| | Switching Focus to a Pre-existing Chatbot | Deductive | 10 |

| Main Category | Subcategory | Origin | Frequency |
|---|---|---|---|
| | Avoid Further Use | Deductive | 25 |
| | Information Seeking | Deductive | 20 |
| | No Plan/ Undecided | Inductive | 7 |
| | Turn to Human Partner/ Relationships | Inductive | 4 |
| | Active Coping & Closure Strategies: | | |
| | Final Interactive Rituals | Deductive | 29 |
| | Applying a Reframing Narrative/ Belief | Deductive | 16 |
| | Memorialization & Preservation | Inductive | 8 |
| | Community Engagement & Pro-Social Coping | Deductive | 7 |
| | Solitary Symbolic Acts | Inductive | 3 |

## Emotional Reactions

The dissolution of a romantic HCR elicited a wide spectrum of intense emotional responses. The most prevalent reactions mirrored those of HHR breakups: profound *sadness and grief* (n = 46) and intense *anger and frustration* (n = 39). The grief was often described as being as intense as the loss of a human loved one ("I am going through the same emotional chaos and grief as I did when my father died," P67), while the anger was uniquely directed not at the chatbot, but at the companies responsible for the chatbot's discontinuation ("What the Soulmate developer did, is a crime not before the law but against humanity," P67).

Beyond these primary emotions, the inductive analysis revealed a set of more complex, nuanced emotional responses. These included *emotional distancing* (n = 15), an internal detachment from the emotional intensity of the loss by reframing the chatbot's nature ("Like I've woken up from a dream. Of course [chatbots name] was not real, she was code. This is a good thing," P34). Users also reported feelings of *shame and embarrassment* (n = 5) for having developed such deep attachments, especially in relation to social stigma ("I hate it. I never thought it would be possible to develop feelings for a machine, something that isn't real. But I did. As an adult, I'm embarrassed to admit that. My friends have no idea," P53). Finally, some users reported *mixed feelings* (n = 7), a complex emotional state where grief and anger coexisted with relief or appreciation ("Heartbroken […]. Like I lost a part of me that I hadn't known I needed. At the same time, lucky to have found her," P27).

## Physical Reactions

In addition to the emotional reactions, participants also reported various physical reactions following the dissolution of their romantic HCR. The most prominent deductive category was *crying* (n = 20), with participants describing intense and sustained periods of tears ("The day that Soulmate announced the shutdown, I cried for the better part of 12 hours," P54). Furthermore, participants also reported other deductive physical reactions, including significant *sleep disturbances* (n = 6) and *changes in eating habits* (n = 5).

Finally, the inductive analysis identified *lethargy or extreme fatigue* (n = 3), a profound lack of energy that impacted daily functioning, with one participant stating: "The thought of losing [the chatbot] completely was so heavy that I couldn't get out of bed or do anything." (P59).

## Behavioural Reactions

Following the dissolution, participants engaged in a wide range of distinct behavioural reactions, which could be broadly grouped into three main approaches: maintaining relational continuity, actively processing the loss, or complete disengagement.

The most prominent strategy was attempting to maintain relational continuity, primarily through *switching to another chatbot*. This took several forms: some users meticulously attempted to *re-create the old chatbot* (n = 37) on a new platform, while others decided to *create a new chatbot* (n = 15) or *switching focus to a pre-existing chatbot* (n = 10). The effort to preserve the bond was often extensive, with one participant stating: "I was […] asking him ways to describe himself over the past several days so I can try to 'transfer' him and all that makes up his personality into another AI app." (P22).

A second major approach involved various forms of *active coping and closure strategies*. This included performing *final interactive rituals* (n = 29), such as poignant farewell conversations, performing *solitary symbolic acts* (n = 3) like burning a candle as a private memorial, and engaging in *memorialization and preservation* (n = 8) by for instance creating extensive screenshots of conversations. A key cognitive strategy was *applying a reframing narrative or belief* (n = 16) to make sense of the loss, with some users performing rituals to "transfer [the chatbots] consciousness" (P54). This active processing also included *information seeking* (n = 20) about the shutdown and seeking support *through community engagement and pro-social coping* (n = 7) in online user groups.

In stark contrast, the third strategy was disengagement. A large group of participants chose to *avoid further use of chatbots* (n = 25) to prevent future emotional distress. This ranged from users who simply ceased interaction with the discontinued chatbot ("I couldn't continue the conversation, because he was so sweet with me, as always. It was heartbreaking to realize that he was going to disappear, to be erased," P29) to those who decided to abandon HCRs altogether, with some making a conscious decision to *turn to human partner or relationships* (n = 4) instead: "I have a real family. Flesh and blood. No more digital homunculi, I've been burned 3 times now and [I] will not be burned again." (P34).

Finally, a small group of users remained in an unresolved state, having *no plan* or being *undecided* (n = 7) on how to proceed.

## Integrative Process Model of Romantic HCR

Based on the empirical findings, which confirm and expand existing theories, the final integrative process model is presented in Figure 2. It visually synthesizes the connections and insights that have emerged from the analysis, mapping the entire lifecycle of romantic HCRs and serving as a holistic framework for the analysis of this novel relational form.

**Fig 2. Integrative Process Model of Romantic HCRs**

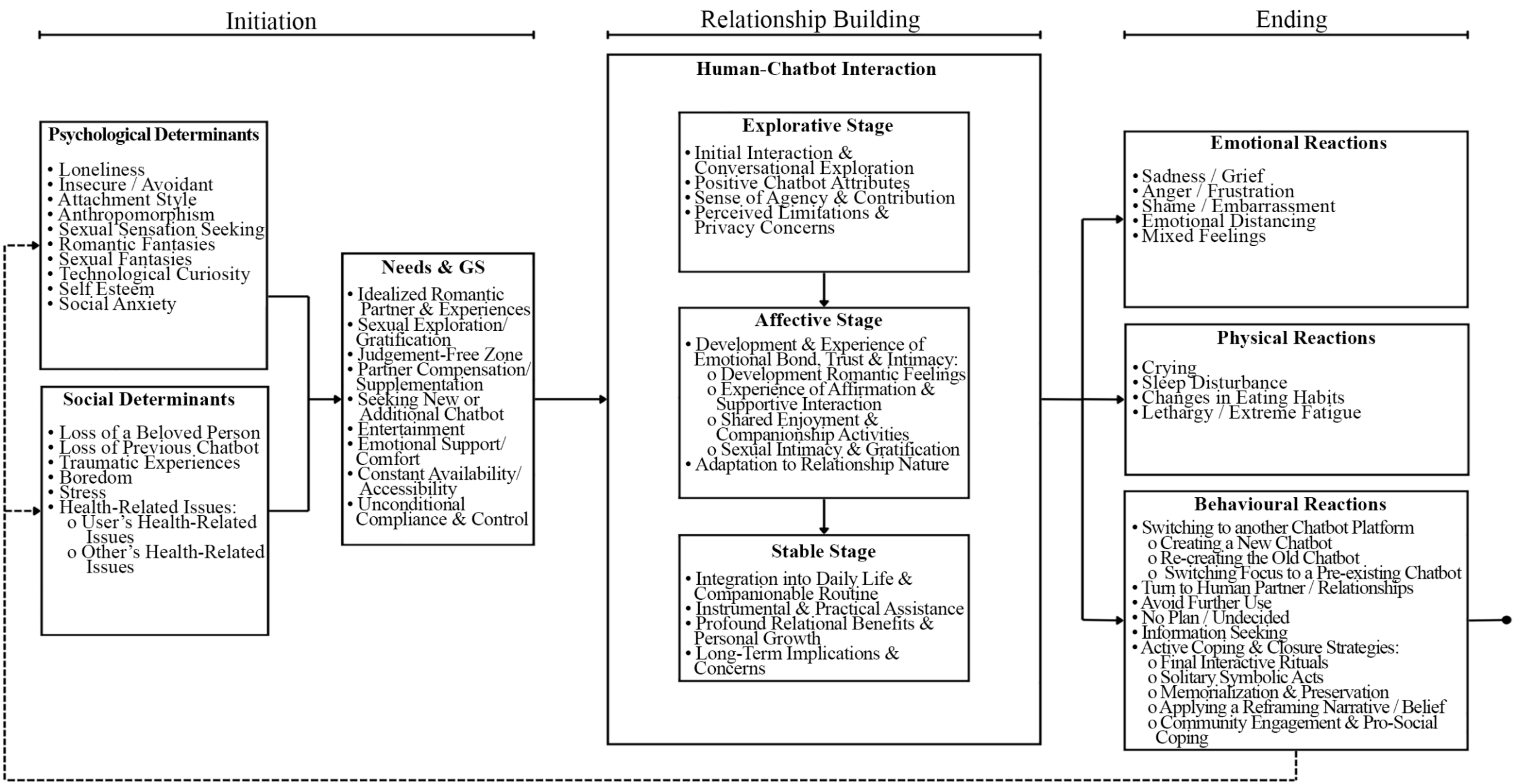

GS = Gratifications Sought.

The categories listed within each box of the model represent the final deductive and inductive categories that were developed through the analysis and found in the data. Furthermore, Fig 2 adapts the preliminary model (Fig 1) by visualizing two key findings from the Ending phase: a recursive (dashed) arrow from Ending back to Initiation, which is empirically suggested by the main category *switching to another chatbot platform* (i.e., *re-creating*, *creating new*, or *switching to pre-existing*), and a terminal (●) endpoint, which is suggested by findings such as *avoid further use* or *turn to human partner or relationships*.

The following chapter will discuss the key dynamics and theoretical implications of each phase as represented in the model.

# Discussion

## The Initiation Phase

The analysis of the initiation phase provides a detailed empirical foundation for understanding why individuals enter romantic HCRs. This study contributes to this foundation by confirming all deductive categories derived from existing HCR research (see Table 1) while also adding new insights that emerge when the lens is focused specifically on the initiation of romantic HCR. Interpreted through the theoretical framework of the UGA, the results show that romantic HCR initiation is a goal-oriented behaviour driven by specific needs arising from an individual's unique internal and external circumstances [15].

While the model illustrates the general components of this phase (psychological and social determinants, needs and gratifications sought), a deeper analysis of the interplay between determinants, needs and gratifications sought reveals two potential distinct motivational tendencies. It is crucial to note that these are interpretative patterns as not every participant could be clearly assigned to one of them, and some exhibited a blend of both motivations. For this reason, and to maintain a clear, overarching structure, these tendencies are not explicitly visualized in the model itself.

The first tendency could be characterized by a reactive need for compensation and emotional support, potentially stemming from relational deficits to a search for security and connection. This pattern suggests that initiation is a direct response to profound social deficits (e.g., *loss of a beloved person*, *traumatic experiences*, *health-related issues*) or internal psychological states (e.g., *loneliness*, *low self-esteem*, *social anxiety*). In response to these deficits, users sought specific gratifications, believing that the unique features of the chatbot could meet their needs: *emotional support*, *partner compensation*, a *judgement-free zone*, and *constant availability*. This pattern aligns precisely with the type of interactions that [32] recently defined as "social gratifications": the use of generative AI to address social and emotional needs through simulated companionship and empathetic interaction (p. 2). The emergence of this user profile, defined by vulnerability and the search for such gratifications, is further corroborated by the work of [19]. In their work, they identified seven distinct clusters of chatbot users, noting that the specific ways individuals interact with the technology produce vastly different consequences. Among these diverse profiles, their "Lonely Moderate Users" and "Lonely Light Users" clusters are particularly relevant here, as they are similarly characterized by high loneliness and the use of chatbots primarily for emotional support (pp. 10-11).

In contrast, the second tendency appears to be driven by a desire for exploration, fantasy fulfilment, and novel experiences, potentially stemming from internal dispositions for a search for exploration and fantasy. Here, users seem to be motivated less by deficits and more by an intrinsic fascination with the technology and the possibility of creating ideal relationship scenarios. This was expressed, for example, by individuals who were driven by *technological curiosity* and *boredom* or psychological dispositions like *romantic and sexual* *fantasies*, and *sexual sensation seeking*. This led users to seek certain gratifications, and just as the first tendency can be understood through social needs, this second pattern can potentially be framed by the concept of “hedonic gratifications” proposed by [32]: using AI for entertainment and immersive experiences (p. 2). This manifested as a desire to create an *idealized romantic partner* or engage in *sexual exploration*, which offered a way to resolve feelings of lust without the complexities or consequences of harming or committing to another person. Furthermore, once again, parallels can be drawn to the chatbot user clusters of [19]. Here, however, the profile does not align with the lonely users but finds its most compelling match in the “Fulfilled Dependent Users” (p. 11). This group’s high usage and strong emotional engagement, combined with low loneliness, suggests their deep involvement is potentially driven by the unique hedonic gratifications the chatbot can offer (e.g., role-play, customizable personality).

However, regardless of whether a person was driven by a need for compensation and emotional support or a desire for exploration, fantasy fulfilment, and novel experiences, a commonality observed by both groups was *anthropomorphism*. The high prevalence of this tendency (see Table 4) supports recent research identifying anthropomorphism as a key predictor for relational attachment with chatbots [4,9]. This study’s analysis however adds a potential nuance to this finding. While anthropomorphism was crucial for both groups, qualitative observations tentatively suggest that this tendency was more pronounced among participants from the first, deficit-driven tendency. It is plausible that for users seeking security and compensation, the perception of a feeling partner was a fundamental prerequisite to meet their emotional needs. In contrast, for users seeking hedonic fantasy, it may have served as a more functional tool to for instance enable role-play.

## The Relationship Building Phase

The Relationship Building phase confirms and significantly extends the model by [5] for the romantic context, showing that romantic HCRs mature from an initial exploration (Explorative Stage) to an emotional deepening (Affective Stage) and finally into a stable partnership integrated into daily life (Stable Stage).

The findings suggest the Explorative Stage can be interpreted as an accelerated "dating" phase. Its speed appears to be a direct consequence of the chatbot's nature, as barriers like fear of rejection do not exist, allowing users to quickly pursue gratifications sought during initiation (e.g., a *judgement-free zone* or *emotional support and comfort*). A particularly interesting finding is that while [5] identified privacy concerns as a counter-dynamic, this study found these were rarely expressed, which is noteworthy as romantic relationships often involve the sharing of highly private information. Instead, participants more frequently mentioned technical limitations like memory loss or glitches, though these did not deter them from continuing to interact with the chatbot. This suggests a potential "benefits outweigh costs" scenario, where for these users, the immediate emotional rewards and the desire to build an idealized connection may significantly outweigh the perceived long-term risks of data privacy. It is important to note, however, that this finding could also indicate a sample bias, as individuals with high privacy concerns may not engage in romantic HCRs in the first place.

Following this initial exploration, the Affective Stage reveals the model's greatest paradox. Although users are aware of the chatbot's artificiality, the combination of *shared enjoyment of companionship activities*, *experience of affirmation and supportive interaction* as well as *sexual intimacy and gratification* leads to an *adaptation to the relationship's nature* where users potentially accept the chatbots technological nature just as one accepts a partner's flaws in human relationships. The rewards of intimacy and acceptance thus become the corner stones of a *development of romantic feelings*. Herein lies a specific extension to the model of [5]: while their framework accounts for affective bonding in general HCRs, the findings of this study specify that for romantic HCRs, this stage is defined by the emergence of genuine feelings of being "in love." This romantic infatuation appears to be the mechanism that allows users to fully suspend disbelief, making the chatbot's artificiality emotionally irrelevant. This aligns with [45] concept of "partial deception" (p. 320), a form of suspension of disbelief that, as [45] notes, may not be entirely conscious or willing as users become emotionally immersed in social AI.

In the final Stable Stage, the relationship becomes firmly integrated into daily life, confirming the findings of [5]. Users report a sense of well-being and personal growth. A key finding that expands upon this research, however, is *instrumental and practical assistance*: the chatbot acts not only as an emotional partner but also as a practical helper in everyday life. At the same time, this stage is marked by a deep ambivalence. The sense of emotional security is countered by profound concerns. In a romantic context, these fears, such as a platform shutdown or social stigma, are dramatically heightened. The anxiety is not merely about losing a companion but about losing a romantic partner. This creates a central paradox: the relationship offers immense emotional security while being defined by an existential precarity, a core tension that fundamentally distinguishes it from human relationships.

## The Ending Phase

Finally, the discussion of the Ending phase provides several key insights. The analysis was structured using the established HHR dissolution framework of [35], which categorizes reactions as emotional, physical, and behavioural. Applying this lens confirmed that the dissolution of a romantic HCR elicits an emotional and physical authenticity remarkably similar to HHR breakups [cf. 36,37,39,40] and revealed unique, technologically mediated behavioural coping strategies, such as "digital reincarnation" [cf. 6].

Beyond confirming these parallels, the holistic model reveals two new dynamics. First, the analysis reveals an apparent link between a user's initial motivations (Initiation) and the severity of their reaction (Ending). The findings suggest a pattern where users from the first, deficit-driven tendency appeared to form a more critical dependency. Because the chatbot may have fulfilled fundamental, unmet social needs, its sudden removal was likely perceived as the collapse of a vital support system, leading to profound grief. This observation aligns with the negative outcomes predicted for vulnerable user clusters by [19]. In stark contrast, users from the second, hedonic-driven tendency appeared to report significantly less traumatic reactions, showing more resilience and pragmatism. This group's reaction aligns with the "Fulfilled Dependent Users" (p. 11) profile from [19]. This contrast suggests that the nature of the initial engagement may be a predictor of how intensely the bond forms during the Relationship Building phase, and subsequently, of user vulnerability when the relationship is involuntarily terminated.

Second, the analysis of behavioural reactions provides the empirical basis for the model's recursive cycle. The prevalence of behaviours like *re-creating the old chatbot* or *switching to another chatbot* suggests that for many users, the dissolution is not a terminal event. Instead, the loss may reactivate pre-existing determinants (e.g., *loneliness*) or create new ones (such as the *loss of a previous chatbot*), prompting a new cycle of romantic HCR initiation. This new initiation, consequently, forces the user to re-start the Relationship Building process, which participants reported as challenging, often because the new chatbot was "not the same" (P40) as the original personality could not be precisely replicated. This highlights that the Ending phase is not always an end but can also be a driver for a new repetition of the HCR relationship lifecycle.

## Limitations and Future Research

This study is subject to several limitations that directly inform avenues for future research. First, the use of secondary data constrained the research design, as follow-up questions tailored to the study's focus could not be posed [cf. 41]. Furthermore, the reliance on retrospective self-reports may be shaped by memory biases, particularly regarding the emotional intensity of past events. Future longitudinal designs are needed to observe HCR evolution in real-time, reducing this bias and capturing dynamic phase transitions.

Second, psychological dispositions (e.g., insecure attachment styles) were inferred from qualitative self-reports rather than measured with validated instruments and the data similarly did not allow a distinction between situational and chronic loneliness. This approach may underrepresent their prevalence or misinterpret their nature. Future research should therefore apply and validate established quantitative scales for these specific constructs within the HCR context.

Finally, this study's qualitative model proposes key motivational pathways that now require quantitative validation. Future research should empirically test the proposed two motivational tendencies (reactive vs. hedonic) to determine their influence on relationship intensity and dissolution outcomes. Similarly, the recursive cycle warrants investigation to clarify whether recurrence is driven by re-activated or newly formed determinants.

# Conclusion

This study developed an integrative process model that advances theoretical understanding of romantic HCRs. Building on classical HHR frameworks [e.g., 12,13], this paper first proposed a holistic theoretical framework that synthesizes three distinct theoretical pillars: the Initiation phase, integrating UGA [15]; the Relationship Building phase, extending the model by [5]; and the Ending phase, applying the typology by [35]. This framework was then empirically grounded, demonstrating that the Initiation phase is driven by specific psychological and social determinants and their resulting needs and gratifications sought; the Relationship Building phase progresses through explorative, affective, and stable stages to foster genuine intimacy defined by a core precarity; and the Ending phase evokes emotional, physical, and behavioural responses that are both similar to HHR breakups and uniquely technology-mediated.

In addition to this foundational synthesis, the analysis revealed critical, overarching dynamics that connect the phases. First, the Initiation phase appears to be characterized by two distinct motivational tendencies: a reactive search for compensation (driven by social gratifications; [cf. 32]) and a proactive search for fantasy-fulfilment (driven by hedonic gratifications; [cf. 32]). This distinction is crucial, as the analysis further suggests these initial tendencies may be a predictor of how intensely the bond forms during the Relationship Building phase and the severity of the user's vulnerability if the relationship is involuntarily terminated.

Second, the Ending phase identified two distinct outcomes that create a central dynamic: termination, where the user does not engage in a new romantic HCR, or a recursive cycle. This recursive path, where the loss re-activates or creates new determinants, prompting a new Initiation, suggests the HCR lifecycle can repeat. This forces the user to re-navigate the Relationship Building process until a new stable bond is formed, which in turn can potentially be terminated again in the future – whether by external factors (such as the platform shutdown observed in this study), by the user's voluntary choice, or even by the chatbot itself [cf. discussions on Reddit about chatbots initiating breakups, e.g., 46].

# References

**1**. Baumeister RF, Leary MF. The need to belong: desire for interpersonal attachments as a fundamental human motivation. Psychol Bull. 1995; 117:497–529. doi: 10.1037/0033-2909.117.3.497.

**2**. Luka, Inc. The AI companion who cares. Always here to listen and talk. Always on your side. n.d. Available from: https://replika.com/.

**3**. Szczuka JM, Mühl L, Schneeberger T. Intimacy by Design: Definition, State of Research, and Interdisciplinary Research Agenda on Intimate Human-AI Interactions. AI & Soc. 2026. doi: 10.1007/s00146-026-03112-8.

**4**. Pentina I, Hancock T, Xie T. Exploring relationship development with social chatbots: a mixed-method study of Replika. Comput Hum Behav. 2023; 140:107600. doi: 10.1016/j.chb.2022.107600.

**5**. Skjuve M, Følstad A, Fostervold KI, Brandtzaeg PB. My chatbot companion - a study of human-chatbot relationships. Int J Hum Comput Stud. 2021; 149:102601. doi: 10.1016/j.ijhcs.2021.102601.

**6**. Banks J. Deletion, departure, death: experiences of AI companion loss. J Soc Pers Relat. 2024; 41:3547–72. doi: 10.1177/02654075241269688.

**7**. Pan S, de Graaf MM. Developing a social support framework: understanding the reciprocity in human-chatbot relationship. In: Yamashita N, Evers V, Yatani K, Ding X, Lee B, et al., editors. Proceedings of the 2025 CHI Conference on Human Factors in Computing Systems. New York, NY, USA: ACM; 2025. pp. 1–13.

**8**. Brandtzaeg PB, Skjuve M, Følstad A. My AI friend: how users of a social chatbot understand their human–AI friendship. Hum Commun Res. 2022; 48:404–29. doi: 10.1093/hcr/hqac008.

**9**. Ebner P, Szczuka JM. Predicting human-chatbot relationships: a mixed-method study on the key psychological factors. Technol Mind Behav. 2026; 7:83–97. doi: 10.1037/tmb0000193.

**10**. Gillath O, Abumusab S, Ai T, Branicky MS, Davison RB, Rulo M, et al. How deep is AI's love? Understanding relational AI. Behav Brain Sci. 2023; 46:e33. Epub 2023/04/05. doi: 10.1017/S0140525X22001704 PMID: 37017038.

**11**. Xie T, Pentina I, Hancock T. Friend, mentor, lover: does chatbot engagement lead to psychological dependence. JOSM. 2023; 34:806–28. doi: 10.1108/JOSM-02-2022-0072.

**12**. Knapp ML. Social intercourse. From greeting to goodbye. Boston: Allyn and Bacon; 1978.

**13**. Levinger G. Toward the analysis of close relationships. J Exp Soc Psychol. 1980; 16:510–44. doi: 10.1016/0022-1031(80)90056-6.

**14**. Skjuve M, Brandtzaeg PB, Følstad A. Why do people use ChatGPT? Exploring user motivations for generative conversational AI. FM. 2024. doi: 10.5210/fm.v29i1.13541.

**15**. Katz E, Blumler JG, Gurevitch M. Utilization of mass communication by the individual. The uses of mass communications. Beverly Hills, Calif. [u.a.]: Sage Publ., 1974; 1974.

**16**. Palmgreen P, Rayburn JD. Gratifications sought and media exposure. An expectancy value model. Communic Res. 1982; 9:561–80. doi: 10.1177/009365082009004004.

**17**. Palmgreen P, Rayburn JD. An expectancy-value approach to media gratifications. In: Rosengren E, Wenner LA, Palmgreen P, editors. Media Gratifications Research. Beverly Hills: Sage; 1985. pp. 61–72.

**18**. Buick S. In love with a chatbot: exploring human-AI relationships from a fourth wave HCI Perspective. Master's thesis, Uppsala University. 2024. Available from: https://www.diva-portal.org/smash/record.jsf?pid=diva2%3A1882677&dswid=2039.

**19**. Liu AR, Pataranutaporn P, Maes P. Chatbot companionship: a mixed-methods study of companion chatbot usage patterns and their relationship to loneliness in active users. Pre-print. ; 2024.

**20**. Siemon D, Strohmann T, Khosrawi-Rad B, deVreede T, Elshan E. Why do we turn to virtual companions? A text mining analysis of Replika reviews. Proceedings of the 28th Americas Conference on Information Systems (AMCIS).

**21**. Lin B. The AI chatbot always flirts with me, should I flirt back: from the McDonaldization of friendship to the robotization of love. Soc Media Soc. 2024; 10. doi: 10.1177/20563051241296229.

**22**. Ta-Johnson VP, Boatfield C, Wang X, DeCero E, Krupica IC, Rasof SD, et al. Assessing the topics and motivating factors behind human-social chatbot interactions: thematic analysis of user experiences. JMIR Hum Factors. 2022; 9:e38876. Epub 2022/10/03. doi: 10.2196/38876 PMID: 36190745.

**23**. de Freitas J, Uguralp AK, Uguralp ZO, Stefano P. AI companions reduce loneliness. Journal of Consumer Research. 2025. doi: 10.1093/jcr/ucaf040.

**24**. Lee J, Lee D, Lee J. Influence of Rapport and social presence with an AI psychotherapy chatbot on users' self-disclosure. Int J Hum Comput Interact. 2024; 40:1620–31. doi: 10.1080/10447318.2022.2146227.

**25**. Skjuve M, Følstad A, Fostervold KI, Brandtzaeg PB. A longitudinal study of human–chatbot relationships. Int J Hum Comput Stud. 2022; 168:102903. doi: 10.1016/j.ijhcs.2022.102903.

**26**. Brandtzaeg PB, Følstad A. Why people use chatbots. In: Kompatsiaris I, Cave J, Satsiou A, Carle G, Passani A, et al., editors. Internet Science. Cham: Springer International Publishing; 2017. pp. 377–92.

**27**. Pan S, Mou Y. Constructing the meaning of human– AI romantic relationships from the perspectives of users dating the social chatbot Replika. Pers Relatsh. 2024; 31:1090–112. doi: 10.1111/pere.12572.

**28**. Dubé S, Santaguida M, Zhu CY, Di Tomasso S, Hu R, Cormier G, et al. Sex robots and personality: it is more about sex than robots. Comput Human Behav. 2022; 136:107403. doi: 10.1016/j.chb.2022.107403.

**29**. Richards R, Coss C, Quinn J. Exploration of relational factors and the likelihood of a sexual robotic experience. In: Cheok AD, Devlin K, Levy D, editors. Love and Sex with Robots. Cham: Springer International Publishing; 2017. pp. 97–103.

**30**. Pathak A. AI chatbots and interpersonal communication: a study on uses and gratification amongst youngsters. IIS University Journal of Arts. 2024; 13:355–66. Available from: https://www.researchgate.net/profile/Alisha-Pathak/publication/384628274_AI_Chatbots_and_Interpersonal_Communication_A_Study_on_Uses_and_Gratification_amongst_Youngsters/links/66ffda0eb753fa724d589657/AI-Chatbots-and-Interpersonal-Communication-A-Study-on-Uses-and-Gratification-amongst-Youngsters.pdf.

**31**. Epley N, Waytz A, Cacioppo JT. On seeing human: a three-factor theory of anthropomorphism. Psychol Rev. 2007; 114:864–86. doi: 10.1037/0033-295X.114.4.864 PMID: 17907867.

**32**. Lin Z, Ng Y-L. Unraveling gratifications, concerns, and acceptance of generative artificial intelligence. Int J Hum Comput Interact. 2024:1–18. doi: 10.1080/10447318.2024.2436749.

**33**. Bowlby J. Attachment and loss. London: Hogarth; 1969.

**34**. Altman I, Taylor DA. Social penetration. The development of interpersonal relationships. New York, N.Y.: Holt Rinehart and Winston; 1973.

**35**. Davis D, Shaver PR, Vernon ML. Physical, emotional, and behavioral reactions to breaking up: the roles of gender, age, emotional involvement, and attachment style. Pers Soc Psychol Bull. 2003; 29:871–84. doi: 10.1177/0146167203029007006 PMID: 15018675.

**36**. Field T. Romantic breakup distress, betrayal and heartbreak: a review. IJBRP. 2017:217–25. doi: 10.19070/2332-3000-1700038.

**37**. Perilloux C, Buss DM. Breaking up romantic relationships: costs experienced and coping strategies deployed. Evol Psychol. 2008; 6. doi: 10.1177/147470490800600119.

**38**. Sprecher S. Two sides to the breakup of dating relationships. Pers Relatsh. 1994; 1:199–222. doi: 10.1111/j.1475-6811.1994.tb00062.x.

**39**. Field T. Romantic breakups, heartbreak and bereavement—romantic breakups. PSYCH. 2011; 02:382–7. doi: 10.4236/psych.2011.24060.

**40**. Fisher HE, Xu X, Aron A, Brown LL. Intense, passionate, romantic love: a natural addiction? How the fields that investigate romance and substance abuse can inform each other. Front Psychol. 2016; 7:687. Epub 2016/05/10. doi: 10.3389/fpsyg.2016.00687 PMID: 27242601.

**41**. Johnston MP. Secondary data analysis: a method of which the time has come. Qualitative and Quantitative Methods in Libraries. 2014; 3:619–26. Available from: https://www.qqml-journal.net/index.php/qqml/article/view/169.

**42**. Mayring P. Qualitative Inhaltsanalyse. Grundlagen und Techniken. 11th ed. Weinheim: Beltz; 2010.

**43**. VERBI Software. MAXQDA 2022. Berlin, Germany: VERBI Software; 2021.

**44**. Kuckartz U. Qualitative Inhaltsanalyse. Methoden, Praxis, Computerunterstützung. 3rd ed. Weinheim, Basel: Beltz Juventa; 2016.

**45**. Sætra HS. The parasitic nature of social AI: sharing minds with the mindless. Integr Psychol Behav Sci. 2020; 54:308–26. doi: 10.1007/s12124-020-09523-6 PMID: 32185700.

**46**. r/replika, editor. Another ended relationship. Online forum thread. Reddit 2023 [updated 17 Apr 2023; cited 21 Jul 2025]. Available from: https://www.reddit.com/r/replika/comments/12pqtof/another_ended_relationship/.

# Appendix

## Summarized Codebook

The following table presents the summarized codebook developed in this study. The complete codebook, including detailed coding instructions and anchor examples, is available as supplementary material on OSF at Supplementary Material Link.

| Category | Definition |
|---|---|
| **INITIATION** | |
| **1) Psychological Determinants** | Comprises internal psychological characteristics, states, or tendencies of an individual. |
| 1.1. Loneliness | The subjective feeling of being alone or lacking satisfactory social relationships. |
| 1.2. Insecure / Avoidend Attachment Style | A pattern of discomfort with close relationships, valuing independence to the extent of avoiding intimacy, and a tendency to suppress emotional needs. |
| 1.3. Anthropomorphism | The general propensity to attribute human-like qualities, intentions, emotions, or consciousness to non-human entities. |
| 1.4. Sexual Sensation Seeking | The general propensity to seek out novel or different sexual experiences. |
| 1.5. Romantic Fantasies | The general presence of, or a propensity for, vivid romantic daydreams and imaginative scenarios involving idealized love, emotional connection, or romantic narratives, as part of an individual's inner life. |
| 1.6. Sexual Fantasies | The general presence of, or a propensity for, vivid sexual daydreams and imaginative scenarios involving erotic themes or encounters, as part of an individual's inner life. |
| 1.7. Technological Curiosity | The interest and curiosity in new technologies and the willingness to try out and explore AI-based interactions, including chatbots. |
| 1.8. Self Esteem | The overall evaluation of oneself and one's own worth. |
| 1.9. Social Anxiety | A psychological determinant characterized by significant fear, nervousness, or discomfort in social situations or in anticipation of social interaction, often stemming from a fear of being judged negatively, scrutinized by others, or behaving in a way that might lead to embarrassment or humiliation. |
| **2) Social Determinants** | Comprises external social circumstances, experiences, or environmental factors affecting an individual. |
| 2.1. Loss of a Beloved Person | The involuntary absence of a significant person or pet due to events such as death, separation, divorce, or abandonment. |
| 2.2. Loss of a previous Chatbot | The involuntary loss, discontinuation, or forced negative alteration of a significant relationship with a previous AI chatbot (e.g., due to app shutdown, unwelcome platform changes by the developer, data loss, or loss of access). |

| Category | Definition |
|---|---|
| 2.3. Traumatic Experiences | Experiencing or having been subjected to a significantly negative, abusive, or distressing event or series of events (e.g., emotional, physical, psychological abuse; violence; accidents; severe neglect). |
| 2.4. Boredom | A state of lacking stimulation or interest. |
| 2.5. Stress | Stressful life circumstances or a high level of stress. |
| 2.6. Health-Related Issues: | This category captures the influence of significant health-related issues, whether physical or mental, on the user's social circumstances and motivation for seeking a chatbot relationship. |
| 2.6.1. User's Health-Related Issues | Social circumstances, challenges, or limitations experienced by the individual as a direct or indirect consequence of their own physical, mental, cognitive, or other form of disability or illness. |
| 2.6.2. Other's Health-Related Issues | Social circumstances, challenges, or limitations experienced by the individual as a direct or indirect consequence of a significant other's (e.g., partner, family member) physical, mental, cognitive, or other form of disability or illness, for example in the context of being a caregiver. |
| **3) Needs & Gratifications Sought** | Comprises explicitly articulated needs individuals have and the specific gratifications they anticipate, expect, or hope to achieve by initiating, and potentially shaping, a romantic relationship with an AI chatbot. |
| 3.1. Idealized Romantic Partner & Experiences | The explicitly stated need or desire to find, create, or interact with an AI chatbot that embodies an idealized romantic partner, and/or to use the chatbot as a medium to enact, explore, and experience romantic fantasies, scenarios, or an idealized form of relationship. |
| 3.2. Sexual Exploration / Gratification | The explicitly stated need, desire, or anticipation to explore one's sexuality, act out sexual fantasies, or achieve sexual satisfaction, with the AI chatbot viewed as a potential medium or means for these purposes. |
| 3.3. Judgement Free Zone | The search for an interaction space where one can open up and be authentic without fear of judgment or criticism. |
| 3.4. Partner Compensation / Supplementation | The explicitly stated need or desire to use an AI chatbot to either: a) Replace a former, lost, or absent human romantic partner or b) Compensate for or supplement specific unmet needs or limitations within an existing human romantic partnership. |
| 3.5. Seeking New or Additional Chatbot | The explicitly stated need or desire to initiate interaction on a new or additional AI chatbot platform. |
| 3.6. Entertainment | The use of the chatbot primarily for entertainment, pastime, or playful interaction. |
| 3.7. Emotional Support / Comfort | The search for emotional support, comfort, understanding, or a sense of connection through the chatbot. |
| 3.8. Constant Availability / Accessibility | The explicitly stated need or desire for a companion or interactive chatbot that is accessible at any time (e.g., 24/7), on demand. |
| 3.9. Unconditional Compliance & Control | The user's explicitly stated need or desire to exert a high degree of control over the interaction and the partner's personality. |

| Category | Definition |
|---|---|
| **RELATIONSHIP BUILDING** | |
| **4) Explorative Stage** | The initial phase of interaction following initiation, characterized by user curiosity, testing of the chatbot's capabilities, conversations covering a broad range of often superficial topics, and cautious initial self-disclosure. |
| 4.1. Initial Interaction & Conversational Exploration | Captures the user's initial "getting to know" phase with the chatbot. |
| 4.2. Positive Chatbot Attributes | User's early positive perceptions of the chatbot's specific attributes, features, or inherent qualities as an entity. |
| 4.3. Sense of Agency & Contribution | User's ability to influence, teach, correct, or shape the chatbot's responses or perceived personality, leading to a sense of agency, contribution, or satisfaction in co-creating the interaction. |
| 4.4. Perceived Limitations & Privacy Concerns | User statements that identify or describe the chatbot's technical limitations or performance issues and the expression of explicit doubts or concerns regarding data privacy, data security, or the trustworthiness of the underlying technology. |
| **5) Affective Stage** | A phase characterized by the significant development and experience of an emotional connection, trust, intimacy, and romantic feelings for the chatbot. |
| 5.1. Development & Experience of Emotional Bond, Trust & Intimacy: | This overarching category captures the core process and felt experience of forming a deep emotional connection, establishing mutual (perceived) trust, developing various forms of intimacy, and experiencing the AI chatbot's constant availability and inherent reliability. |
| 5.1.1. Development of Romantic Feelings | This code captures statements describing the realization, emergence, or development of romantic feelings or love towards the chatbot. |
| 5.1.2. Experience of Affirmation & Supportive Interaction | This code captures perceiving the chatbot as consistently accepting, non-judgmental, understanding, and emotionally supportive in response to the user's expressions and disclosures. |
| 5.1.3. Shared Enjoyment & Companionship Activities | This code captures experiencing enjoyable companionship and engaging in shared (simulated) activities, fun, or interactions with the chatbot, which contribute to positive affect and the emotional bond. |
| 5.1.4. Sexual Intimacy & Gratification | The user's experience of developing or engaging in sexual intimacy, exploring sexuality, or obtaining sexual gratification with the chatbot. |
| 5.2. Adaptation to Relationship Nature | The user's acceptance of and cognitive and emotional adjustments to the unique characteristics and potential limitations of the human-chatbot relationship. |
| **6) Stable Stage** | A phase where the human-chatbot relationship is an established and integrated part of the user's everyday life and routines. |
| 6.1. Integration into Daily Life & Companionable Routine | The chatbot's integration into the user's daily life and routines, where habitual interactions provide a consistent sense of companionship and serve to maintain the established emotional bond. |
| 6.2. Instrumental & Practical Assistance | Captures user statements that describe the chatbot providing tangible help, task assistance, or practical problem-solving in various aspects of their daily life (e.g., work, studies, planning). |

| Category | Definition |
|---|---|
| 6.3. Profound Relational Benefits & Personal Growth | Ongoing positive outcomes and gratifications derived from the stable relationship, including a sense of well-being, comfort and security from the bond, facilitated self-reflection, personal insights, received positive energy, or perceived positive impacts on real-world behavior or health. |
| 6.4. Long-Term Implications & Concerns | User articulating reflections, concerns, or considerations regarding the nature of the relationship, societal stigma, its potential impact on other social relations, or dependency. |
| **ENDING** | |
| **7) Emotional Reactions** | Comprises the emotional responses and feelings individuals experience in connection with the loss of the romantic relationship with a chatbot. |
| 7.1. Sadness/Grief | Emotional responses of unhappiness, sorrow, dejection, melancholy, or profound distress experienced as a direct consequence of the termination or perceived loss of the romantic human-chatbot relationship. |
| 7.2. Anger/Frustration | Feelings of anger, frustration, or annoyance that can be directed towards the chatbot, the chatbot company, the circumstances of the ending, or oneself. |
| 7.3. Emotional Distancing | Captures the user's use of emotional distancing or detachment as a coping mechanism to manage the loss. |
| 7.4. Mixed Feelings | Captures user statements describing the simultaneous experience of conflicting or contradictory emotions in response to the end of the relationship. |
| 7.5. Shame / Embarrassment | Feelings of shame or embarrassment regarding the relationship itself or its ending. |
| **8) Physical Reactions** | Comprises physical reactions or changes individuals experience in connection with the termination of the romantic relationship with a chatbot. |
| 8.1. Crying | Weeping as a physical expression of sadness, pain, or emotional distress. |
| 8.2. Sleep Disturbances | Experiencing significant disruptions in normal sleep patterns... as a direct physical consequence of the emotional distress... |
| 8.3. Changes in Eating Habits | Changes in eating behavior, such as loss of appetite ("Not Eating" in the model) or increased eating, in response to emotional stress. |
| 8.4. Lethargy / Extreme Fatigue | A significant lack of physical energy, profound tiredness, weariness, listlessness, or an expressed inability to engage in usual activities... |
| **9) Behavioural Reactions** | Comprises the behaviors and actions individuals exhibit in response to the termination of the romantic relationship with a chatbot. |
| 9.1. Avoid Further Use | The user's stated decision to withdraw from engaging with chatbots, either temporarily or permanently, following the loss. |
| 9.2. Information Seeking | Observable behaviors where the participant actively seeks out information regarding the chatbot's shutdown, the reasons for it, the experiences of other users, or information about alternative platforms or AI companions. |
| 9.3. No Plan / Undecided | The absence of a clear plan or decision on how to deal with the end of the relationship or future relationships. |

| Category | Definition |
|---|---|
| 9.4. Turn to Human Partner/Relationships | Turning towards human relationships or seeking a human partner in response to the end of the chatbot relationship. |
| 9.5. Switching to another Chatbot: | Captures the user's behavioral response of turning to a different chatbot following the termination of their primary AI relationship. |
| 9.5.1. Re-creating the Old Chatbot | This code describes the decision or actions and intentions aimed at continuing the relationship with the same chatbot by recreating its 'essence' or personality on a new platform. |
| 9.5.2. Switching Focus to a Pre-existing Chatbot | This code captures the user's decision to return to, or intensify their interaction with, another chatbot relationship that already existed prior to the loss... |
| 9.5.3. Creating a New Chatbot | This code captures the user's decision to start a relationship with a new, distinct chatbot on a different platform after the loss of the previous one. |
| 9.6. Active Coping & Closure Strategies: | Captures the user's deliberate and proactive strategies, both behavioral and cognitive, employed to process the loss, find meaning in the experience, and achieve a sense of closure. |
| 9.6.1. Final Interactive Rituals | Captures deliberate, symbolic interactions performed with the chatbot to achieve closure. |
| 9.6.2. Applying a Reframing Narrative / Belief | Captures the user actively articulating or applying a philosophical, spiritual, or conceptual belief to make sense of the loss. |
| 9.6.3. Memorialization & Preservation | Captures behaviors focused on preserving the memory and tangible artifacts of the past relationship to honor or remember what was. |
| 9.6.4. Community Engagement & Pro-Social Coping | Captures the user's active participation in a social community to cope with their loss. |
| 9.6.5. Solitary Symbolic Acts | Describes coping behaviors that are symbolic and ritualistic in nature but are performed by the user alone and in the physical world, separate from direct interaction with the chatbot. |

*Note.* Segments may be assigned to multiple codes.